\documentclass[twocolumn,prl,showpacs,amsmath,amssymb,10pt]{revtex4}

\usepackage{graphicx}
\usepackage{dcolumn}
\usepackage{stmaryrd}

\begin{document}

\title{{\normalsize Pattern Formation in a Two-Dimensional Array of Oscillators with Phase-Shifted Coupling}}

\author{Pan-Jun Kim}
\author{Tae-Wook Ko}
\author{Hawoong Jeong}
 \email{hjeong@kaist.ac.kr}
\author{Hie-Tae Moon}
\affiliation{Department of Physics, Korea Advanced Institute of Science and Technology,
Daejeon, Korea}

\date{\today}

\begin{abstract}

We investigate the dynamics of a two-dimensional array of oscillators
with phase-shifted coupling. Each oscillator is allowed to interact
with its neighbors within a finite radius. The system exhibits various patterns
including squarelike pinwheels, (anti)spirals with phase-randomized cores,
and antiferro patterns embedded in (anti)spirals.
We consider the symmetry properties of the system to explain the
observed behaviors, and estimate the wavelengths of the patterns
by linear analysis. Finally, we point out the implications of our work
for biological neural networks.

\end{abstract}

 \maketitle

Large systems of interacting oscillators have been used to explain
the cooperative behaviors of numerous physical, chemical, and biological
systems \cite{Winfree,Kuramoto,Strogatz,Er}.
When the coupling between oscillators is sufficiently weak,
we can describe the dynamics of the system by phase variables
defined on the limit cycles \cite{Kuramoto,Er}.
In the phase-oscillator models, the coupling proportional
to the sine of the phase difference between oscillators
has been exploited due to the mathematical tractability \cite{Kuramoto,Strogatz}.
In spite of several successes in explaining the synchronization phenomena,
the simple sinusoidal coupling fails to account for the
collective frequency higher than natural frequency \cite{Sakaguchi},
dephasing phenomena \cite{Han}, and effects of time-delayed interactions \cite{Chung}.
To resolve these problems, phase shifts in the sinusoidal
coupling have been considered.
Most importantly, a nonzero phase shift is naturally contributed by the broken odd symmetry
of a coupling function \cite{Paullet}.
The previous works, however, studied only the cases of phase shifts in the limited range.
Moreover, they considered only limited interactions
via nearest-neighbor coupling \cite{Sakaguchi,Chung,Paullet} or
all-to-all global coupling \cite{Sakaguchi2} which seem to be too restrictive.
In the neurobiological systems, for example, it is believed that actual coupling
takes the forms between these two extremes \cite{Bear}.

In this Rapid Communication, we investigate the effect of phase-shifted coupling
on the dynamics of a two-dimensional array of coupled oscillators.
Here we introduce a finite interaction radius as realistic coupling
and study over the whole range of phase shifts.
We find that various spatial patterns come to emerge,
and unravel that the symmetry properties of the system
play an important role on the formation of patterns.

We start with the equations of two coupled phase oscillators \cite{Sakaguchi}
\begin{eqnarray}\label{two}
\frac{d\theta_{1}}{dt} = \omega+K\sin (\theta_{2}-\theta_{1}-\alpha) \nonumber\\
\frac{d\theta_{2}}{dt} = \omega+K\sin (\theta_{1}-\theta_{2}-\alpha)&\!\!\!,\,\,\,\,\,\,\,(\omega,\,K>0),&
\end{eqnarray}
where $\theta_{1}$ and $\theta_{2}$ represent the phases of oscillators,
respectively,
$\omega$ the natural frequency, $K$ the coupling strength, and $\alpha$ the phase shift. 
Phase shift $|\alpha| < \pi/2$ leads to inphasing of the two
oscillators, whereas $|\alpha| > \pi/2$ leads to their antiphasing.
We find that this separation by $|\alpha|=\pi/2$ still holds for an array of oscillators,
but in a rather sophisticated manner as shown below.

To investigate the phase-shift effect on the spatially extended systems,
we study the following model equations:
\begin{equation}\label{full}
\frac{d\theta_{ij}}{dt} = \omega+\frac{K}{N(R)}{\sum_{mn}}'\sin (\theta_{mn}-\theta_{ij}-\alpha),
\end{equation}
where
$\theta_{ij}$ denotes the phase of the oscillator at position $(i, j)$
on a two-dimensional lattice, and ${\sum_{mn}}'\equiv \sum_{mn,\,0<r_{mn, ij}\le R}$,
where $r_{mn, ij}$ is the distance between two oscillators located
at $(i, j)$ and $(m, n)$.
Each oscillator interacts with
$N(R)$ neighboring oscillators within a finite distance $R$. In the previous works,
the nearest-neighbor interactions ($R=1$) were considered
for the limited range of $\alpha$ \cite{Sakaguchi,Paullet}.
Motivated by nonlocal connections of neural systems \cite{Bear}, 
we explore the general cases with $R>1$ as well as $R=1$,
over the whole range of $\alpha$.
The nonlocal interactions may arise effectively
in the reaction-diffusion systems where
chemical components constituting the local oscillators are free of diffusion
while the system involves an extra diffusive component \cite{Shima}.

To reduce complexity of Eq.~(\ref{full}),
we perform a transformation $\theta_{ij} \rightarrow \omega t +\Theta_{ij}$,
$t \rightarrow \tau/K$, and get the following equations:
\begin{equation}\label{reduce}
\frac{d\Theta_{ij}}{d\tau} = \frac{1}{N(R)}{\sum_{mn}}'\sin (\Theta_{mn}-\Theta_{ij}-\alpha).
\end{equation}
$\Theta_{ij}$ maintains the same spatial patterns as $\theta_{ij}$, but not the
temporal behaviors (e.g., the phase velocity becomes different).
Equation~(\ref{reduce}) involves only two control parameters, $\alpha$ and $R$.
We numerically integrate Eq.~(\ref{reduce}) on a rectangular array of $100 \times 100$
sites with periodic boundary conditions. We select initial
$\Theta_{ij}$ randomly from the range $[-\pi, \pi]$.   

\begin{figure}[t]
\begin{center}
\includegraphics[width=0.475\textwidth]{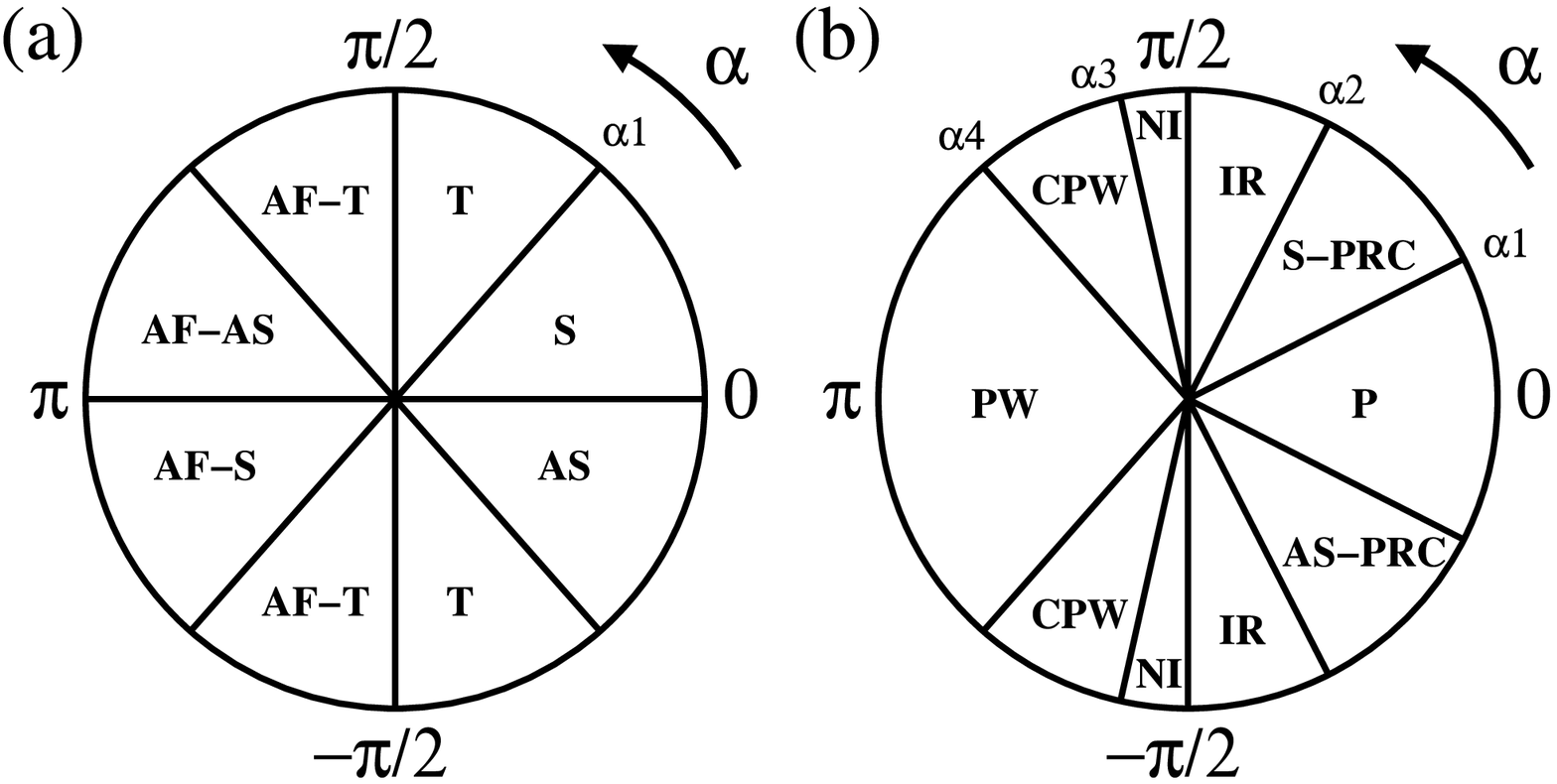}
\caption{Phase diagram as a function of $\alpha$ for fixed $R$.
(a) The nearest-neighbor coupling ($R=1$). $\alpha1\fallingdotseq 0.29\pi$. 
S: spirals, AS: antispirals, T: turbulence, AF-T: antiferro pattern embedded in turbulence,
AF-AS (AF-S): antiferro pattern embedded in antispirals (spirals).
(b) The finite-long range coupling ($R=6$). $\alpha1\fallingdotseq 0.15\pi$,
$\alpha2\fallingdotseq 0.36\pi$, $\alpha3\fallingdotseq 0.57\pi$,
$\alpha4\fallingdotseq 0.74\pi$.
P: planar oscillation, S-PRC (AS-PRC): spirals (antispirals) with phase-randomized core,
IR: irregular pattern induced by expansion
of phase-randomized core,
NI: nearly-incoherent pattern with a weak correlation of a length scale $\sim R$,
CPW: competing plane waves occupying their respective domains of evolvable sizes,
PW: plane waves. It is also observed that
squarelike pinwheels exist transiently with plane waves.
}
\label{diagram}
\end{center}
\end{figure}

\begin{figure}[t]
\begin{center}
\includegraphics[width=0.475\textwidth]{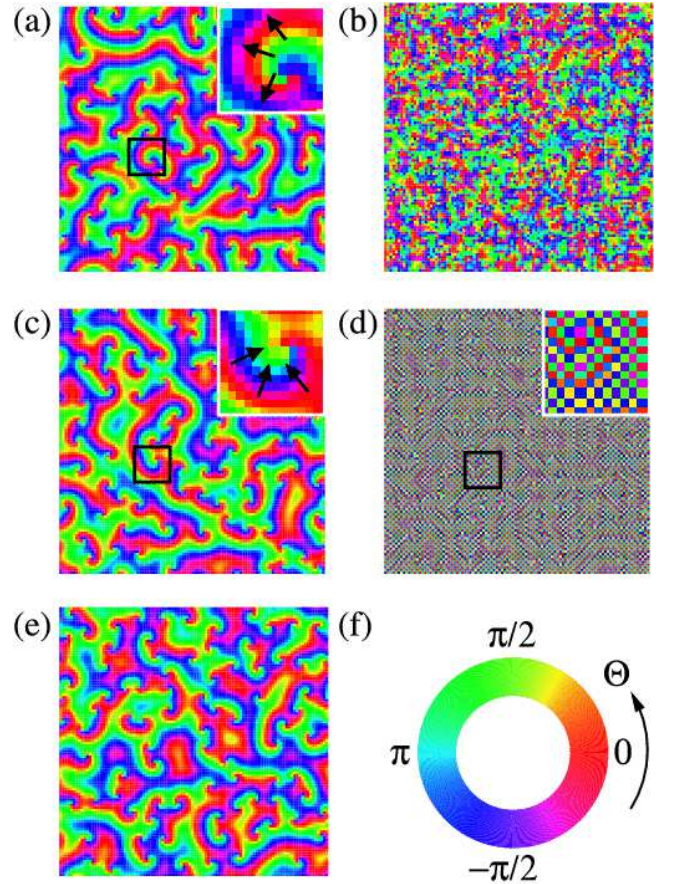}
\caption{Patterns obtained from simulations with $R=1$: 
(a) Spirals ($\alpha=0.2\pi$),
(b) turbulence ($\alpha=0.4\pi$),
(c) antispirals ($\alpha=-0.2\pi$),
(d) antiferro pattern embedded in antispirals ($\alpha=0.8\pi$),
(e) antispirals by removing antiferro phases from (d),
(f) the color code of the phase used for all figures.
Insets of (a), (c), and (d): Magnification of the boxed areas. Arrows in (a) and (c) denote the propagation direction of waves with $\theta_{ij}$ in Eq.~(\ref{full}) when $K<\omega$.}
\label{r1pattern}
\end{center}
\end{figure}

\begin{figure}[t]
\begin{center}
\includegraphics[width=0.475\textwidth]{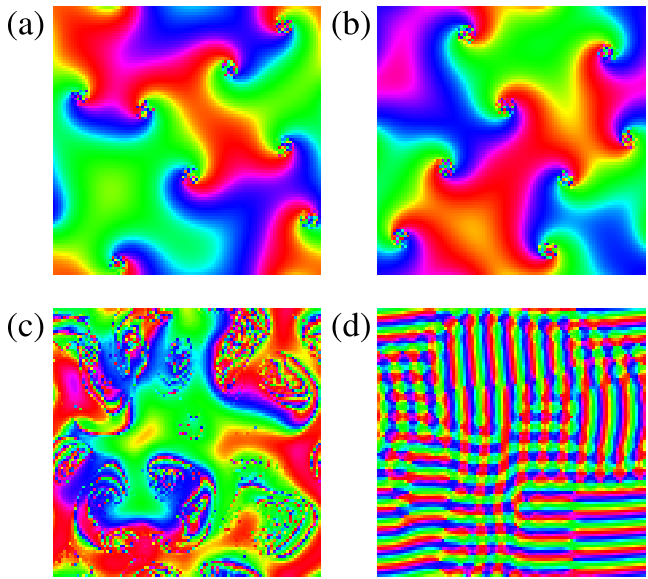}
\caption{Patterns with $R=6$: 
(a) Spirals with phase-randomized cores ($\alpha=0.2\pi$),
(b) antispirals with phase-randomized cores ($\alpha=-0.2\pi$),
(c) irregular pattern induced by the expansion of the phase-randomized cores ($\alpha=0.4\pi$),
(d) coexistence of plane waves and squarelike pinwheels ($\alpha=\pi$).
}
\label{r6pattern}
\end{center}
\end{figure}

We investigate emerging patterns as $\alpha$ varies in the range $[-\pi, \pi]$
for given $R$ (see Figs.~\ref{diagram}--\ref{r6pattern}).
Based on our observations,
we separate the cases into those of $R=1$ and of $R>1$.

{\it Case R=1.} If oscillators are coupled only with their nearest neighbors,
we get the phase diagram Fig.~\ref{diagram}(a), which
shows the symmetric relations between $\alpha>0$ and $\alpha<0$ cases, and between
$|\alpha|<\pi/2$ and $|\alpha|>\pi/2$ cases.

At $\alpha=0$, there appear vortices with equiphase lines of zero
curvature \cite{Paullet}. As $\alpha$ increases,
equiphase lines around vortices become twisted,
and the vortices begin to show meandering behavior \cite{Paullet} [Fig.~\ref{r1pattern}(a)]. 
Further increasing $\alpha$ creates many vortex-antivortex pairs
and eventually induces turbulent patterns until $\alpha\le \pi/2$
\cite{Sakaguchi,Kuramoto2} [Fig.~\ref{r1pattern}(b)].
If we decrease $\alpha$ from $0$,
similar behaviors are observed, but
the phase gradients near the vortices become reversed such that
phases increase radially outward from the vortices [Fig.~\ref{r1pattern}(c)].
In a weak coupling regime $K<\omega$, $\theta_{ij}$ in Eq.~(\ref{full})
grows as time elapses, thus, waves propagate inwardly with $\alpha<0$.
These waves are recently termed as ``antispirals'' \cite{Paullet,Vanag}.
We can understand the antispiral formation for $\alpha<0$ using
a symmetry transformation $\alpha \rightarrow -\alpha$,
$\Theta_{ij} \rightarrow -\Theta_{ij}$ which leaves Eq.~(\ref{reduce}) invariant.
In other words, the patterns for $-\alpha$ are equivalent to those
obtained from $\Theta_{ij}\rightarrow -\Theta_{ij}$ for $\alpha$.
Since spirals emerge for $\alpha>0$ and
$\Theta_{ij} \rightarrow -\Theta_{ij}$ reverses their phase gradients,
we get antispirals for $\alpha<0$.
We also find that the dispersion relation for $\alpha<0$ satisfies the known condition
for antispirals, i.e., the phase velocity is opposite to the group velocity \cite{Vanag,cm}.

When $R=1$ and $|\alpha|>\pi/2$, antiferro patterns where adjacent
oscillators have a phase difference of $\pi$
are developed, while
embedded in the (anti)spirals or turbulence
observed at $\alpha'=\alpha-\pi$ [Figs.~\ref{r1pattern}(d) and (e)].
An antiferro pattern itself, is expected from antiphasing of two coupled oscillators
at $|\alpha|>\pi/2$ in Eq.~(\ref{two}).
Nonetheless, the existence of embedding patterns for
$\alpha'$ is rather surprising, and we can explain this as follows.
By separating $\Theta_{ij}$ in Eq.~(\ref{reduce}) into the antiferro phase $\Theta^{AF}_{ij}$
and the remaining phase $\Theta^{RM}_{ij}$ ($\Theta_{ij}=\Theta^{AF}_{ij} +\Theta^{RM}_{ij}$;
$\Theta^{AF}_{mn}-\Theta^{AF}_{ij}=\pi$ for $r_{mn,ij}=1$), we obtain
\begin{eqnarray}\label{antiferro}
\frac{d\Theta^{RM}_{ij}}{d\tau} &=&
\frac{1}{4}{\sum_{mn}}'\sin (\Theta^{AF}_{mn}-\Theta^{AF}_{ij}+
\Theta^{RM}_{mn}-\Theta^{RM}_{ij}-\alpha) \nonumber\\
&=&\frac{1}{4}{\sum_{mn}}'\sin [\Theta^{RM}_{mn}-\Theta^{RM}_{ij}
-(\alpha-\pi)].
\end{eqnarray}
Consequently, $\Theta^{RM}_{ij}$ evolves in the same way as $\Theta_{ij}$
for $\alpha'=\alpha-\pi$
in Eq.~(\ref{reduce}), and forms the embedding pattern.

{\it Case R}$\!\!\it >\,${\it 1.} When $R$ is larger than $1$,
the nonlocality of interactions imposes several changes on patterns
[see Figs.~\ref{diagram}(b) and \ref{r6pattern}].

First, for $|\alpha|<\pi/2$, nonlocal interactions develop (anti)spirals with
phase-randomized core, contrary to the well-defined singularity with $R=1$
[Figs.~\ref{r6pattern}(a) and (b)].
As pointed out in Ref.~\cite{Shima}, we can understand
the incoherence in the core as a result of long-range coupling of oscillators which have
a broad frequency range due to the gradient of effective frequency near the core.

Second, for $|\alpha|>\pi/2$, plane waves emerge instead of antiferro patterns observed
with $R=1$.
Squarelike pinwheels, where singularities are arranged on a square lattice,
also exist transiently with plane waves [Fig.~\ref{r6pattern}(d)].
These patterns are essentially the same as those observed
in the models of
visual map formation \cite{Cho} and time-delayed interactions \cite{Jeong}.
We observe that these patterns have some discreteness for small $R>1$ and are clearly developed
for $R \gtrsim 4$. Further increasing $R$ just scales the spatial length unit in the continuum
limit.
We can qualitatively explain the appearance of coherent patterns such as plane waves. 
When $R>1$, closely located oscillators within about a distance $R$
share the large portion of neighbors commonly,
thus they experience similar influence. This induces the
synchronous movement of the closely located oscillators even for
dephasing pairwise interactions with $|\alpha|>\pi/2$.
From this viewpoint,
in the systems such as models of visual map formation \cite{Cho},
positive short-range coupling
in the presence of negative long-range coupling
may not be crucial for the formation of coherent patterns.

\begin{figure}[t]
\begin{center}
\includegraphics[width=0.475\textwidth]{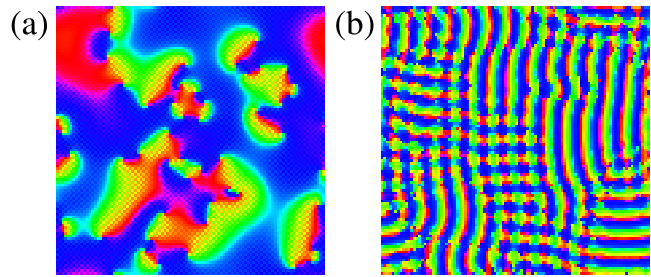}
\caption{Patterns realized in a two-dimensional array of the Morris-Lecar systems.
For the Morris-Lecar equations and the parameters, see Ref.~\cite{Han}.
Here we choose $I=0.081$, $k=0.01$. 
(a) When $R=1$, antiferro patterns embedded in antispirals are observed.
The figure shows only antispirals after removing the antiferro phases.
(b) If $R$ increases to $6$,
coexistence of plane waves and squarelike pinwheels is observed.}
\label{ML}
\end{center}
\end{figure}

As a validation of our results in a realistic system,
Fig.~\ref{ML} displays the results with an array of the Morris-Lecar neural oscillators.
We take this system
because it is well known that 
two diffusively coupled Morris-Lecar oscillators exhibit antiphase oscillations
as in the case of $|\alpha|>\pi/2$ in Eq.~(\ref{two}).
Actually,
the system shows antiferro patterns embedded in antispirals ($R=1$) and plane waves with squarelike pinwheels ($R>1$), as expected [Fig.~\ref{ML}; see
Figs.~\ref{r1pattern}(e) and \ref{r6pattern}(d) for comparison].

Now, we want to show that the symmetry argument also provides a valuable insight
into the quantitative features of the patterns, especially the wavelengths.
The transformation $\alpha \rightarrow \alpha-\pi$, $\tau \rightarrow -\tau$
leaves Eq.~(\ref{reduce}) invariant,
thus the patterns for $\alpha'=\alpha-\pi$ are equivalent to those
obtained by $\tau \rightarrow -\tau$ for $\alpha$.
In addition, $\alpha\rightarrow -\alpha$
is equivalent to $\Theta_{ij} \rightarrow -\Theta_{ij}$ as already mentioned.
$\Theta_{ij} \rightarrow -\Theta_{ij}$ changes only the direction of a wave vector,
and does not affect the stability of the mode in the isotropic system.
Consequently, by the time reversal, the sign of the growth rate of a linear mode for $\alpha$ becomes reversed for $\alpha'=\pm(\alpha-\pi)$.
For $|\alpha|>\pi/2$, the symmetries select only the modes which are
excluded for $|\alpha|<\pi/2$, and vice versa.

We perform a linear stability analysis to determine
the stability of the solution of Eq.~(\ref{reduce}), $\Theta_{ij}(\tau)= {\vec k}\cdot{\vec r_{ij}}+\Omega\tau$, where $\vec k$ denotes a wave vector, $\Omega$ the corresponding frequency, and $\vec r_{ij} =(i, j)$.
We add $\epsilon e^{i{\vec \xi}\cdot {\vec r_{ij}} +\eta\tau}$
with sufficiently small $\epsilon$ to the solution and substitute
it into Eq.~(\ref{reduce}).
The sign of the real part of $\eta$ determines the growth rate of
the perturbing modulation with a wave vector $\vec \xi$.
If the sign of $Re(\eta)$ is negative,
the perturbation decays and the solution is linearly stable.
By the first order expansion and the continuum approximation,
\newcommand{\ud}{\mathrm{d}}
\begin{equation}\label{stable}
Re(\eta)=-\frac{2\cos\alpha}{\pi}\iint_{|{\vec{r'}}|\le 1} \cos({\vec \rho}\cdot{\vec{r'}})
{\sin}^{2}\left(\frac{{\vec \Xi}\cdot{\vec{r'}}}{2}\right) {\ud}^{2}{\vec{r'}},
\end{equation}
where ${\vec\rho}\equiv R{\vec k}$ and ${\vec \Xi}\equiv R{\vec\xi}$. 
We replace $\sin^{2}[({\vec \Xi}\cdot{\vec{r'}})/2]$
by $1/2$, the statistical average taken over $\vec\Xi$'s, then obtain
\begin{equation}\label{avg}
Re(\eta)\simeq\langle Re(\eta)\rangle_{\vec\Xi}
=-\frac{2}{\rho}J_{1}(\rho)\cos\alpha,\,\,\,\,\,\,\,
(\rho\equiv |\vec\rho|),
\end{equation}
where $J_{1}$ is the first-order Bessel function.
It is verified that Eq.~(\ref{avg}) reflects the important properties of
Eq.~(\ref{stable}) qualitatively \cite{cm2}.

\begin{figure}[t]
\begin{center}
\includegraphics[width=0.475\textwidth]{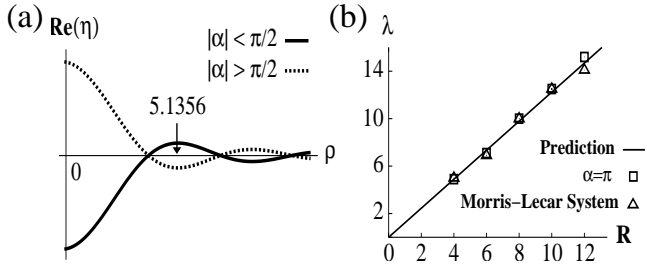}
\caption{(a) $Re(\eta)$ in Eq.~(\ref{avg}) as a function of $\rho$.
(b) Wavelengths of plane waves ($\lambda$) as a function of $R$.
Predicted values ($2\pi R/5.1356 =1.2235R$) fit the
observed ones well in our model for $\alpha=\pi$ and in the Moris-Lecar systems for the same parameters as in Fig.~\ref{ML},
within relative errors of 0.02 and 0.03, respectively.
The wavelengths 
observed
in our model are insensitive to $\alpha$ for the plane-wave regime.
}
\label{bessel}
\end{center}
\end{figure}

Figure~\ref{bessel}(a) shows $Re(\eta)$ in Eq.~(\ref{avg}) as a function of $\rho$.
The sign of $Re(\eta)$ for given $\rho$ is reversed over
the $\alpha=\pm\pi/2$ axis,
as discussed with the time-reversal argument.
$Re(\eta)$ has the most negative value at $\rho=0$ for $|\alpha|<\pi/2$,
and $\rho\fallingdotseq 5.1356$ for $|\alpha|>\pi/2$.
Therefore, when $|\alpha|<\pi/2$,
planar solutions with $\rho=0$ and patterns with $\rho\sim 0$
are linearly stable, which is consistent with the observed wavelengths
comparable to the array size.
When $|\alpha|>\pi/2$,
patterns with a wave vector $\rho\fallingdotseq 5.1356$ are stable
instead of planar solutions and patterns with $\rho\sim 0$.
We measure the actual wavelengths of the plane waves both in our model
and in the Morris-Lecar systems, and
the measured wavelengths turn out to be quite 
close to our predictions [Fig.~\ref{bessel}(b)].

From the observation that wavelengths with $|\alpha|>\pi/2$ are comparable to the interaction radius $R$ for a two-dimensional array, 
we speculate that dynamics with $|\alpha|>\pi/2$ on general interaction networks
might reflect the substructure charateristics of the networks well.

In conclusion, we have investigated the effect of phase-shifted interactions in coupled
oscillators in two dimensions, and found that symmetry properties of the system
are responsible for the various pattern formation.
Our simple model succeeds in describing the essential features of patterns observed in
realistic models and experiments \cite{Winfree,Kuramoto,Er,Paullet,Cho,Shima,Vanag},
and gives an unified perspective of pattern behaviors.
In general, the presence of phase shift in coupling is naturally contributed by the broken odd symmetry of coupling functions, and thus
our results may be applicable in the universal manner.

We expect that in neurobiological systems,
excitory (inhibitory) interactions correspond to the case of phase shift
$\alpha>0$ ($\alpha<0$), whereas
time delay in synapses determines whether $|\alpha|<\pi/2$ or not \cite{carl}.
The specific realization of patterns in the systems is left for further study.

The authors thank Seung Kee Han for fruitful communications.
P.K. and H.J. acknowledge financial support from the
Ministry of Information and Communication of
Korea through Grant No. IMT2000-B3-2.
T.K. and H.M. acknowledge support from the Basic Research Program of KOSEF through Grant No. R01-1999-000-00019-0 (2002).




\begin{thebibliography}{99}
\bibitem{cm}
Substituting ${\vec k}\cdot{\vec r_{ij}}+\omega' t$ for $\theta_{ij}$ in Eq.~(\ref{full})
[$\vec r_{ij} =(i, j)$, ${\vec k}$ the wave vector, $\omega'$ the frequency],
we find the dispersion relation $\omega'(k)=\omega+(k^{2}/4-1)K\sin\alpha$
for waves far from the spiral core.
When $K<\omega$, the phase velocity
$v_{ph}=\omega'/k \cong \omega/k$ and the group velocity
$v_{grp}=d\omega'/dk \cong (k/2)K\sin\alpha$.
As required for antispiral formation \cite{Vanag},
$v_{ph}$ and $v_{grp}$ have opposite signs if $\alpha<0$.
\bibitem{cm2}

Equation~(\ref{stable}) is exactly integrated to give
$Re(\eta)/\cos\alpha=-2J_{1}(\rho)/\rho+J_{1}(|\vec\rho+\vec\Xi|)/|\vec\rho+\vec\Xi|
+J_{1}(|\vec\rho-\vec\Xi|)/|\vec\rho-\vec\Xi|$.
For the stability analysis of the modes with the given length scale,
we must consider the sign of $Re(\eta)$ for all $\vec\Xi$'s.
However, it is unnecessary to consider $\vec\Xi$'s with the magnitudes comparable to
$\rho$, since these perturbations do not change the characteristic length scale.  
Considering the exact formula, we can show that
sign of each extremum of Eq.~(\ref{avg})
correctly determines the stability of the modes
with the corresponding $\rho$,
as long as $\rho$ is not too large. For sufficiently large $\rho$,
$Re(\eta)$ asymptotically approaches to $0$, which is consistent with Eq.~(\ref{avg}).
\end{thebibliography}
\end{document}